\documentclass[twocolumn,superscriptaddress,preprintnumbers,amsmath,amssymb,floatfix,nofootinbib]{revtex4}

\usepackage{graphicx}
\usepackage{dcolumn}
\usepackage{bm}
\begin{document}

\preprint{APS/123-QED}

\title{Mono-, bi- and tri- graphene layers deposited on conductive Au/Cr/SiO$_{2}$/Si substrate studied by scanning tunnelling spectroscopy }

\author{Z. Klusek}
 \email{zbklusek@mvii.uni.lodz.pl}
\affiliation{%
Division of Physics and Technology of Nanometer Structures, Department of Solid State Physics, University of Lodz, 90-236 Lodz, Pomorska 149/153, Poland\\
}%
\author{P.Dabrowski}%
\affiliation{%
Division of Physics and Technology of Nanometer Structures, Department of Solid State Physics, University of Lodz, 90-236 Lodz, Pomorska 149/153, Poland\\
}%

\author{P.J. Kowalczyk}%
\affiliation{%
Division of Physics and Technology of Nanometer Structures, Department of Solid State Physics, University of Lodz, 90-236 Lodz, Pomorska 149/153, Poland\\
}%
\author{W. Kozlowski}%
\affiliation{%
Division of Physics and Technology of Nanometer Structures, Department of Solid State Physics, University of Lodz, 90-236 Lodz, Pomorska 149/153, Poland\\
}%
\author{P. Blake}
\affiliation{
Graphene Industries Ltd. Manchester Centre for Mesoscience and Nanotechnology,
University of Manchester, Oxford Road, Manchester M13 9PL, UK
}%
\author{M. Szybowicz}%
\affiliation{%
Faculty of Technical Physics, Poznan University of Technology, Nieszawska 13A, 60-965 Poznan, Poland
\\
}%
\author{T. Runka}%
\affiliation{%
Faculty of Technical Physics, Poznan University of Technology, Nieszawska 13A, 60-965 Poznan, Poland
\\
}%
\author{W. Olejniczak}%
\affiliation{%
Division of Physics and Technology of Nanometer Structures, Department of Solid State Physics, University of Lodz, 90-236 Lodz, Pomorska 149/153, Poland\\
}%
\date{\today}

\begin{abstract}
Graphene devices require electrical contacts with metals, particularly with gold. Scanning tunneling spectroscopy studies of electron local density of states performed on mono-, bi- and tri- graphene layer deposited on metallic conductive Au/Cr/SiO$_2$/Si substrate shows that gold substrate causes the Fermi level shift downwards which means that holes are donated by metal substrate to graphene which becomes p-type doped. These experimental results are in good accordance with recently published density function theory calculations. The estimated positions of the Dirac point show that the higher number of graphene layers the lower Fermi level shift is observed.
\end{abstract}

\maketitle

Recently, it has been shown that graphene possesses unique electronic properties due to the mass-less Dirac fermion character of carriers derived from the conical dispersion relation close to the Dirac point \cite{novo1,novo2}. This leads to the presence of many physical effects in this material, namely unconventional quantum Hall effect \cite{novo2,zha3}, ballistic transport of electrons \cite{berg4}, electronic spin transport \cite{tom5}, micron scale coherence length \cite{mor6}, or single electron tunneling \cite{wes7}. Unusual physical properties have been also reported on bilayer and biased bilayer graphene \cite{oht8,nil9}.  It seems to be crucial for future applications to understand the nanoscale electronic properties of graphene considered in terms of the electron local density of states (LDOS) on different substrates. This is because a substrate can affect considerably the electronic properties of graphene i.e. change the position of the Fermi level (E$_F$) relative to the Dirac point (E$_D$). These nanoscale studies have been carried out by scanning tunneling microscopy/spectroscopy techniques (STM/STS) on graphene lying on highly oriented pyrolitic graphite (HOPG) \cite{li10},  graphene/SiC \cite{rut11,rut12,bra13,lau14,vaz15}, and mechanically cleaved graphene on SiO$_2$/Si with tunable back-gate electrodes \cite{zha16}. Since electronic transport measurements in graphene devices require metallic contacts, it is tempting to know the physical properties of graphene/metal interface in nanoscale \cite{ya17,sch18,uch19,gio20}. Recently it has been proved theoretically (density functional theory - DFT) that the electronic structure of graphene around K point is preserved on (111) surfaces of Al, Cu, Ag, Pt, and Au \cite{gio20}.  This is especially important in the case of Au which is widely used in fabrication of metal-graphene contacts. Furthermore, these metal substrates cause the Fermi level shift downwards which means that holes are donated by metal substrates to graphene which becomes p-type doped i.e. E$_F$ is located below E$_D$.

In this letter we report on studies of the LDOS of mono-, bi- and tri- graphene layer (MG, BG, TG) deposited on metallic conductive Au/Cr/SiO$_2$/Si substrate. This type of substrate enables us to create a setup suitable for the STM/STS experiments without micro fabrication processes and studies of the LDOS on multilayer graphene systems. We demonstrate that Raman spectroscopy (RS) which has been recently used to identify  MG, BG and TG \cite{fer21,cal22}, shows the potential to evaluate a number of graphene layers on conductive Au/Cr/SiO$_2$/Si substrate. Furthermore, it is commonly known that the three interface Fresnel-law-based model is effective to explain the optical contrast of graphene on SiO$_2$/Si \cite{bla23}. By adding a 4$^{th}$ interface to the model it was proved that sufficiently thin metallic layers deposited onto oxidized silicon before flake preparation do not obscure the optical interference effect that makes graphene visible. This enables graphene experiments that require electrically conductive substrates such as STM/STS with instantaneous optical inspection. Finally, we present our  STS conductance maps which show a distinct electronic contrast between MG, BG and TG leading to identification of the number of graphene layers. A detailed analysis of STS spectra shows the position of the Dirac point above the Fermi level as it is expected from the density functional theory calculations of graphene on gold.

Our sample was prepared by 8 nm of Au, with 0.5 nm Cr adhesion layer sputtered onto 90 nm SiO$_2$. The Au layer is thick enough to be continuous, but still sufficiently thin to maintain optical contrast. The image of flakes on gold in white light and with 530 nm light shows the enhanced contrast when using narrow band interference filters. The Raman investigations were performed in back-scattering geometry using a Renishaw InVia Raman microscope equipped with a confocal DM 2500 Leica optical microscope, a thermoelectrically (TE)-cooled RenCam CCD detector and an Ar$^+$ ion laser working at a wavelength  of
 488 nm. The applied laser power before focusing with 100x objective magnification was less than 0.5 mW.
An edge filter was used to reject Rayleigh scattering. The Raman spectra were carried out in the 1400-3000 cm$^{-1}$ spectral range and they were accumulated 20 times to obtain a better signal to noise ratio. All the STM/STS experiments were carried out at room temperature using a VT-STM/AFM integrated with the MULTIPROBE P UHV system (Omicron GmbH). In the STS mode the I/V curves were recorded simultaneously with a constant current image by the use of an interrupted-feedback-loop technique. Based on these measurements the first derivative of the tunnelling current with respect to the voltage (dI/dV) being a measure of the LDOS was numerically calculated and used to build spatial conductance maps i.e. dI/dV(x,y,E).
\begin{figure}
\includegraphics{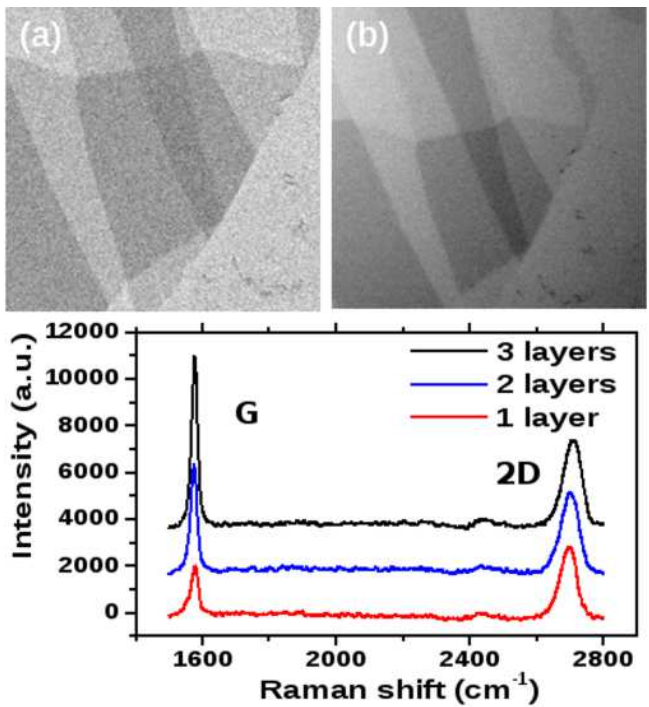}
\caption{\label{fig1}(color online). (a),(b) OM (left) and SEM (right) images of MG, BG and TG deposited onto Au/Cr/SiO2/Si substrate. Typical Raman spectra of MG, BG and TG are presented below.}
\end{figure}

In Figure 1(a) and (b) we present optical microscopy (OM) and scanning electron microscopy (SEM) images of MG, BG and TG onto conductive Au/Cr/SiO$_2$/Si substrate. In the case of optical imaging, a narrow band pass filter at 530 nm  was used to enhance the contrast. From the presented results it is clear that 8 nm of Au with 0.5 nm Cr adhesion layer onto 90 nm SiO$_2$ still allows sufficient optical contrast for graphene identification. Typical Raman spectra recorded on MG, BG and TG are also presented in Fig.1. The detailed analysis of the RS spectra shows a single component of the 2D peak at 2696 cm$^{-1}$ on the graphene regions. This value is different in comparison with graphene on other substrates such as SiO$_2$, SiC, GaAs, glass or graphite \cite{yin24}. The Raman spectra recorded on BG show four components of the 2D peak at 2657 cm$^{-1}$, 2678 cm$^{-1}$, 2697 cm$^{-1}$ and 2718 cm$^{-1}$. Similar measurements on TG show two components of the 2D peak at 2697 cm$^{-1}$ and 2723 cm$^{-1}$ \cite{ram25}.
\begin{figure}
\includegraphics{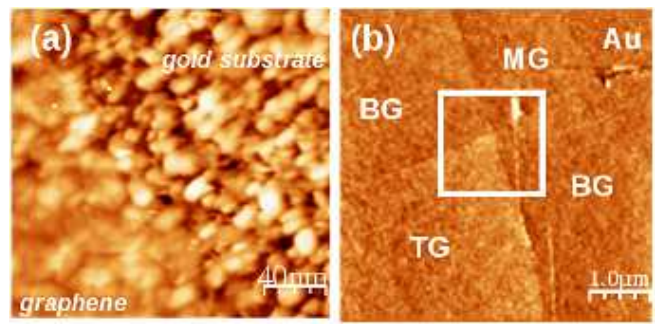}
\caption{\label{fig2}(color online). (a) The 200 nm x 200 nm STM topography showing gold-graphene interface. (b) The 5444 nm x 5444 nm STM topography showing coexistence of MG, BG and TG flakes. STS measurements were carried out in the surface region denoted by a square.}
\end{figure}

In Fig.2(a) we present 200 nm x 200 nm STM topography of MG/gold interface recorded at sample bias U=+0.8 V and the tunneling current set point equals 0.2 nA. From the cross-section profiles we estimated that the height of  graphene on the gold substrate equals roughly 0.5 nm. Furthermore, the graphene layer does not  identically map the structure of the gold substrate - estimated value of RMS calculated over 200 nm x 200 nm area on graphene equals 0.25 nm while a similar measurement on gold gives value close to 0.65 nm. The STM topography (sample bias U=+0.8 V, the current set point 0.2 nA) of Au, MG, BG and TG which coexists together is presented in Fig.2(b). Crucial for our studies is the fact that graphene flakes of different thickness were investigated simultaneously using the tunnelling tip having the same electronic structure. This is especially important when STS measurements in the region denoted by a square in Fig.2(b) were carried out. The height of BG relative to the monolayer  varies from  0.34 nm up to 0.50 nm. Similar measurements for TG  relative to BG give the value in the range of  0.35 -0.50 nm. The estimated value of RMS calculated over 200 nm x 200 nm area  is very similar for BG and TG roughly equals 0.25 nm.
\begin{figure}
\includegraphics{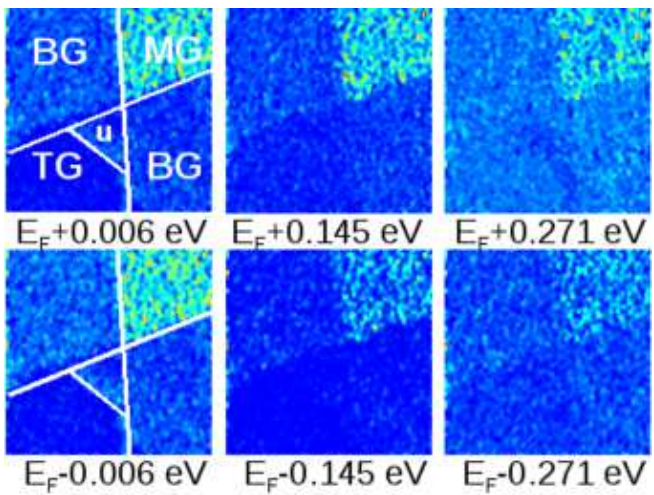}%
\caption{\label{fig3}(color online: blue, green, red - low, intermediate, and high value of the LDOS, respectively). 500 nm x 500 nm dI/dV(E,x,y) maps recorded on MG, BG and TG}
\end{figure}

In Fig.3 dI/dV(x,y,E) maps at selected energies around the Fermi level on mono-, bi- and tri-layer of graphene  are shown. It is clear that MG on Au has very high LDOS values close to the Fermi level (E$_F$ $^+$/$_-$0.006 eV). Furthermore, we observe that the spatial distribution of the LDOS amplitudes measured at the same energy is not uniform and varies from point to point. It simply leads to the conclusion that graphene is not homogenous in terms of local electronic structure. Much lower LDOS values were demonstrated on bilayer graphene, and especially on trilayer. The observed electronic contrast between MG, BG and TG was observed without difficulty at low energies around the Fermi level. However, at energies higher than E$_F$ $^+$/$_-$0.5 eV, the spatial distribution of the LDOS is rather homogenous and it is difficult to distinguish between MG, BG and TG. The result showing that MG on Au has much higher LDOS close to E$_{F}$ than BG and TG seems to be intriguing. This is because we expect higher LDOS in multilayer graphene compared to that for monolayer graphene. However, we would like to emphasize that even though the electronic structure of graphene around K point is preserved, we still observe a strong influence of metallic gold character on the measured LDOS. This influence seems to be much less in the case of BG and TG  as it is observed in our experiments. In the region denoted by ‘u’ in Fig.3 we observed a mixture of the electronic structure typical for TG and BG deposited on gold.

In Fig.4 we present dI/dV(E,line) maps calculated as a function of energy and position along selected  lines on MG, BG and TG regions (TG’ – the region with peculiar electronic structure) and appropriate cross section profiles i.e. dI/dV(E). In all cases we observe that the conductance map shows asymmetry between occupied and unoccupied states i.e. the LDOS is much higher at occupied states for all energies. In the case of MG at the unoccupied part of the spectra, we observe a very pronounced maximum (max) of the LDOS which spreads over 13 nm. On the map, energy of the state varies from 0.29 eV up to 0.38 eV above the Fermi level. It means that MG is not homogeneous and its electronic structure depends strongly on the position on the surface. Furthermore, we show a representative dI/dV curve taken at the position denoted by a horizontal line on the dI/dV(E,line) map. The dI/dV curve shows a very pronounced maximum (max) located at energy close to 0.35 eV above the Fermi level accompanied with the local minimum at 0.39 eV making the spectrum asymmetric around E$_F$. In our interpretation, the minimum marks the position of the Dirac point E$_D$ on monolayer graphene deposited on gold. Taking into account 65535 individual dI/dV curves we estimated that the position of  E$_D$ is located in the range of  0.35 - 0.40 eV above the Fermi level. From the DFT calculations of the Fermi level shift ($\Delta$E$_F$) as a function of the graphene-metal surface distance the 0.35 eV - 0.4 eV shift takes place when a graphene-Au gap is in the range of 0.38 - 0.40 nm \cite{gio20}. This theoretical value is very close to our height measurements of MG on gold substrate which equal 0.5 nm. The small discrepancy can be caused by the fact that STM does not show structure of the surface in crystallographic sense; hence estimation of the height of the observed MG on gold is strongly affected by the electronic structure of the investigated surface and the tunnelling tip. Furthermore, we observe a few (1,2,3) local maxima on the spectrum having different amplitudes. We believe that these local maxima are related to complicated electronic structure of MG/Au system.
\begin{figure}
\includegraphics{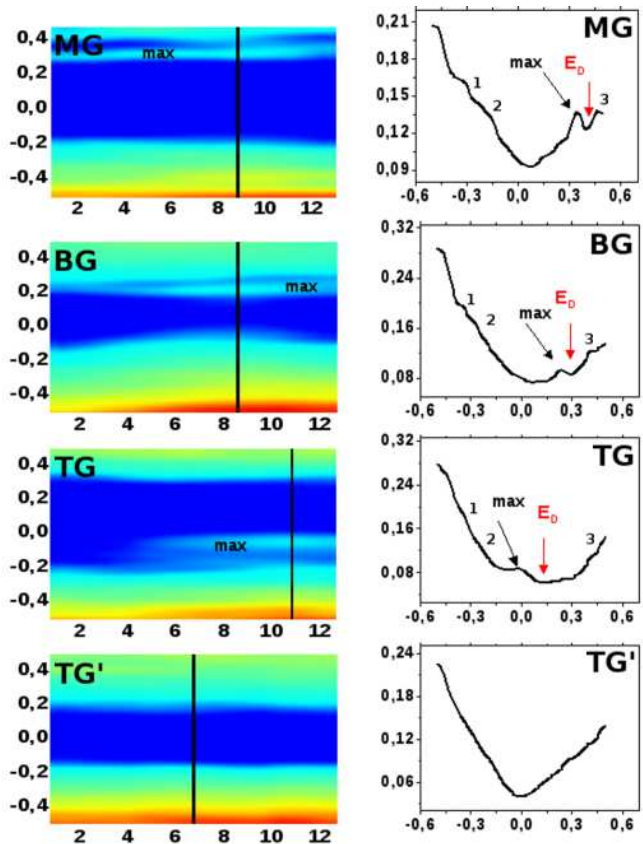}%
\caption{\label{fig4}(color online): blue, green, red - low, intermediate, and high value of dI/dV, respectively). Left column. dI/dV(E,line) maps recorded on MG, BG and TG/TG’. Vertical scale - energy in eV relative to the Fermi level, horizontal scale - length in nanometers.  Right column. dI/dV profiles. Vertical scale – dI/dV, horizontal scale energy in eV relative to the Fermi level}
\end{figure}
Similar results were collected in the case of BG - see the conductance map and representative dI/dV curve. In this case the position of E$_D$ varies from 0.22 eV up to 0.30 eV, leading to the conclusion that the Fermi level shift for BG is less than for MG. In our opinion these results show that the Dirac point for MG and BG deposited on gold substrate is located above the Fermi level as it is expected for hole doping \cite{gio20}. We emphasize that local maxima (1,2,3) in the case of BG are much weaker in comparison with MG, which can be interpreted in terms of a screening effect.

Intriguing conductance maps and dI/dV curves were collected on trilayer graphene as it is presented in Fig.3 TG and TG’. Firstly, we observed a region of trilayer grahene (Fig.3TG) in which the conductance map shows a very pronounced local maximum (max) of the LDOS just below the Fermi level. This maximum is accompanied with the broad local minimum centered at 0.14 eV above the Fermi level making the spectrum enormously asymmetric in the vicinity of E$_F$. We believe that the minimum at 0.14 eV marks the position of E$_D$ on trilayer graphene deposited on gold and shows that the Fermi level shift for TG is less than for MG and BG. Secondly, we observe a peculiar region on trilayer grahene (TG’) in which the conductance map is featureless and the typical dI/dV curve shows parabolic shape having only local maxima with extremely low amplitudes. What is more, in the small energy range ($^+$/$_-$50 meV) the dI/dV curve is symmetric around E$_F$. A change of the electrical properties from the TG to TG’ type takes place over 25-50 nm which suggests that we are not dealing with an abrupt transformation. The parabolic character of the dI/dV spectrum can be explained bearing in mind continuous transformation of trilayer graphene into a fourlayer system. Recently, it has been proved by angle-resolved photoemission spectroscopy (ARPES) that the shift of the Fermi level decreases considerably with a number of graphene layers \cite{oht26}. In the case of TG’ region we can simply assume that $\Delta$E$_F$=0 and the Dirac point is exactly centered at the Fermi level.

In conclusion, we have used OM, SEM, RS and STM/STS techniques to identify and study the electronic structure of MG, BG and TG deposited on conductive Au/Cr/SiO$_2$/Si substrate. We demonstrate that a thin, conductive Au layer on SiO$_2$ still allows sufficient optical contrast for MG, BG and TG identification. The STM results show that the height of MG relative to Au substrate is close to 0.5 nm, the height of BG relative to MG varies from 0.34 nm up to 0.50 nm, while TG relative to BG gives the value 0.35 - 0.50 nm. The STS results prove that holes are donated by Au substrate to graphene which becomes p-type doped i.e. E$_F$ is located below E$_D$. Estimated positions of the Dirac point show that the higher number of graphene layers the lower Fermi level shift is observed. The estimation of the position of E$_D$ in the case of MG on Au is in good accordance with recently published DFT calculations.\\

\begin
{acknowledgments}
The assistance of L. Kowalczyk during paper preparation is appreciated.
\end{acknowledgments}
$^*$This work was financially supported by Lodz University grant UL NR. 505/694.
\bibliography{grapheneAuCrSiO2}

\begin{thebibliography}{26}
\expandafter\ifx\csname natexlab\endcsname\relax\def\natexlab#1{#1}\fi
\expandafter\ifx\csname bibnamefont\endcsname\relax
  \def\bibnamefont#1{#1}\fi
\expandafter\ifx\csname bibfnamefont\endcsname\relax
  \def\bibfnamefont#1{#1}\fi
\expandafter\ifx\csname citenamefont\endcsname\relax
  \def\citenamefont#1{#1}\fi
\expandafter\ifx\csname url\endcsname\relax
  \def\url#1{\texttt{#1}}\fi
\expandafter\ifx\csname urlprefix\endcsname\relax\def\urlprefix{URL }\fi
\providecommand{\bibinfo}[2]{#2}
\providecommand{\eprint}[2][]{\url{#2}}

\bibitem[{\citenamefont{Novoselov et~al.}(2004)}]{novo1}
\bibinfo{author}{\bibfnamefont{K.~S.} \bibnamefont{Novoselov}}
  \bibnamefont{et~al.}, \bibinfo{journal}{Science}
  \textbf{\bibinfo{volume}{306}}, \bibinfo{pages}{666} (\bibinfo{year}{2004}).

\bibitem[{\citenamefont{Novoselov et~al.}(2005)}]{novo2}
\bibinfo{author}{\bibfnamefont{K.~S.} \bibnamefont{Novoselov}}
  \bibnamefont{et~al.}, \bibinfo{journal}{Nature (London)}
  \textbf{\bibinfo{volume}{438}}, \bibinfo{pages}{197} (\bibinfo{year}{2005}).

\bibitem[{\citenamefont{Zhang et~al.}(2005)}]{zha3}
\bibinfo{author}{\bibfnamefont{Y.}~\bibnamefont{Zhang}} \bibnamefont{et~al.},
  \bibinfo{journal}{Nature (London)} \textbf{\bibinfo{volume}{438}},
  \bibinfo{pages}{201} (\bibinfo{year}{2005}).

\bibitem[{\citenamefont{Berger et~al.}(2006)}]{berg4}
\bibinfo{author}{\bibfnamefont{C.}~\bibnamefont{Berger}} \bibnamefont{et~al.},
  \bibinfo{journal}{Science} \textbf{\bibinfo{volume}{312}},
  \bibinfo{pages}{1191} (\bibinfo{year}{2006}).

\bibitem[{\citenamefont{Tombros et~al.}(2007)}]{tom5}
\bibinfo{author}{\bibfnamefont{N.}~\bibnamefont{Tombros}} \bibnamefont{et~al.},
  \bibinfo{journal}{Nature} \textbf{\bibinfo{volume}{448}},
  \bibinfo{pages}{571} (\bibinfo{year}{2007}).

\bibitem[{\citenamefont{Morozov et~al.}(2005)}]{mor6}
\bibinfo{author}{\bibfnamefont{S.~V.} \bibnamefont{Morozov}}
  \bibnamefont{et~al.}, \bibinfo{journal}{Phys.\ Rev.\ B}
  \textbf{\bibinfo{volume}{72}}, \bibinfo{pages}{201401}
  (\bibinfo{year}{2005}).

\bibitem[{\citenamefont{Westervelt}(2008)}]{wes7}
\bibinfo{author}{\bibfnamefont{R.}~\bibnamefont{Westervelt}},
  \bibinfo{journal}{Science} \textbf{\bibinfo{volume}{320}},
  \bibinfo{pages}{324} (\bibinfo{year}{2008}).

\bibitem[{\citenamefont{Ohta et~al.}(2006)}]{oht8}
\bibinfo{author}{\bibfnamefont{T.}~\bibnamefont{Ohta}} \bibnamefont{et~al.},
  \bibinfo{journal}{Science} \textbf{\bibinfo{volume}{313}},
  \bibinfo{pages}{951} (\bibinfo{year}{2006}).

\bibitem[{\citenamefont{Nilsson et~al.}(2008)}]{nil9}
\bibinfo{author}{\bibfnamefont{J.}~\bibnamefont{Nilsson}} \bibnamefont{et~al.},
  \bibinfo{journal}{Phys.\ Rev.\ B} \textbf{\bibinfo{volume}{78}},
  \bibinfo{pages}{045405} (\bibinfo{year}{2008}).

\bibitem[{\citenamefont{Li et~al.}(2222)}]{li10}
\bibinfo{author}{\bibfnamefont{G.}~\bibnamefont{Li}} \bibnamefont{et~al.},
  \bibinfo{journal}{arXiv} \textbf{\bibinfo{volume}{0803}},
  \bibinfo{pages}{4016} (\bibinfo{year}{2222}).

\bibitem[{\citenamefont{Rutter et~al.}(2007{\natexlab{a}})\citenamefont{Rutter,
   et~al.}}]{rut11}
\bibinfo{author}{\bibfnamefont{G.}~\bibnamefont{Rutter}}, ,
  \bibnamefont{et~al.}, \bibinfo{journal}{Science}
  \textbf{\bibinfo{volume}{317}}, \bibinfo{pages}{219}
  (\bibinfo{year}{2007}{\natexlab{a}}).

\bibitem[{\citenamefont{Rutter et~al.}(2007{\natexlab{b}})}]{rut12}
\bibinfo{author}{\bibfnamefont{G.~M.} \bibnamefont{Rutter}}
  \bibnamefont{et~al.}, \bibinfo{journal}{Phys.\ Rev.\ B}
  \textbf{\bibinfo{volume}{76}}, \bibinfo{pages}{235416}
  (\bibinfo{year}{2007}{\natexlab{b}}).

\bibitem[{\citenamefont{Brar et~al.}(2007)}]{bra13}
\bibinfo{author}{\bibfnamefont{V.~W.} \bibnamefont{Brar}} \bibnamefont{et~al.},
  \bibinfo{journal}{App.\ Phys.\ Lett.} \textbf{\bibinfo{volume}{91}},
  \bibinfo{pages}{1222102} (\bibinfo{year}{2007}).

\bibitem[{\citenamefont{Lauffer et~al.}(2008)}]{lau14}
\bibinfo{author}{\bibfnamefont{P.}~\bibnamefont{Lauffer}} \bibnamefont{et~al.},
  \bibinfo{journal}{Phys.\ Rev.\ B} \textbf{\bibinfo{volume}{77}},
  \bibinfo{pages}{155426} (\bibinfo{year}{2008}).

\bibitem[{\citenamefont{de~Parga et~al.}(2008)}]{vaz15}
\bibinfo{author}{\bibfnamefont{A.~L.~V.} \bibnamefont{de~Parga}}
  \bibnamefont{et~al.}, \bibinfo{journal}{Phys.\ Rev.\ Lett.}
  \textbf{\bibinfo{volume}{100}}, \bibinfo{pages}{056807}
  (\bibinfo{year}{2008}).

\bibitem[{\citenamefont{Zhang et~al.}(2008)}]{zha16}
\bibinfo{author}{\bibfnamefont{Y.}~\bibnamefont{Zhang}} \bibnamefont{et~al.},
  \bibinfo{journal}{Nature\ Phys.} \textbf{\bibinfo{volume}{4}},
  \bibinfo{pages}{627} (\bibinfo{year}{2008}).

\bibitem[{\citenamefont{Blanter et~al.}(2007)}]{ya17}
\bibinfo{author}{\bibfnamefont{Y.~M.} \bibnamefont{Blanter}}
  \bibnamefont{et~al.}, \bibinfo{journal}{Phys.\ Rev.\ B}
  \textbf{\bibinfo{volume}{76}}, \bibinfo{pages}{155433}
  (\bibinfo{year}{2007}).

\bibitem[{\citenamefont{Schomerus}(2007)}]{sch18}
\bibinfo{author}{\bibfnamefont{H.}~\bibnamefont{Schomerus}},
  \bibinfo{journal}{Phys.\ Rev.\ B} \textbf{\bibinfo{volume}{76}},
  \bibinfo{pages}{045433} (\bibinfo{year}{2007}).

\bibitem[{\citenamefont{Uchoa et~al.}(2008)}]{uch19}
\bibinfo{author}{\bibfnamefont{B.}~\bibnamefont{Uchoa}} \bibnamefont{et~al.},
  \bibinfo{journal}{Phys.\ Rev.\ B} \textbf{\bibinfo{volume}{77}},
  \bibinfo{pages}{035420} (\bibinfo{year}{2008}).

\bibitem[{\citenamefont{Giovannetti et~al.}(2008)}]{gio20}
\bibinfo{author}{\bibfnamefont{G.}~\bibnamefont{Giovannetti}}
  \bibnamefont{et~al.}, \bibinfo{journal}{Phys.\ Rev.\ Lett.}
  \textbf{\bibinfo{volume}{101}}, \bibinfo{pages}{026803}
  (\bibinfo{year}{2008}).

\bibitem[{\citenamefont{Ferrari et~al.}(2006)}]{fer21}
\bibinfo{author}{\bibfnamefont{A.}~\bibnamefont{Ferrari}} \bibnamefont{et~al.},
  \bibinfo{journal}{Phys.\ Rev.\ Lett.} \textbf{\bibinfo{volume}{97}},
  \bibinfo{pages}{187401} (\bibinfo{year}{2006}).

\bibitem[{\citenamefont{Calizo et~al.}(2008)}]{cal22}
\bibinfo{author}{\bibfnamefont{I.}~\bibnamefont{Calizo}} \bibnamefont{et~al.},
  \bibinfo{journal}{J. Phys. Conf. Ser.} \textbf{\bibinfo{volume}{109}},
  \bibinfo{pages}{012008} (\bibinfo{year}{2008}).

\bibitem[{\citenamefont{Blake et~al.}(2007)}]{bla23}
\bibinfo{author}{\bibfnamefont{P.}~\bibnamefont{Blake}} \bibnamefont{et~al.},
  \bibinfo{journal}{App.\ Phys.\ Lett.} \textbf{\bibinfo{volume}{91}},
  \bibinfo{pages}{063124} (\bibinfo{year}{2007}).

\bibitem[{\citenamefont{ying Wang et~al.}(2008)}]{yin24}
\bibinfo{author}{\bibfnamefont{Y.}~\bibnamefont{ying Wang}}
  \bibnamefont{et~al.}, \bibinfo{journal}{J. Phys. Chem. C}
  \textbf{\bibinfo{volume}{112}}, \bibinfo{pages}{10637}
  (\bibinfo{year}{2008}).

\bibitem[{ram(2008)}]{ram25}
\bibinfo{journal}{The detailed Raman spectroscopy studies of MG, BG and TG on
  Au/Cr/SiO2/Si substrate will be presented in separate paper which is in
  preparation}  (\bibinfo{year}{2008}).

\bibitem[{\citenamefont{Ohta et~al.}(2007)}]{oht26}
\bibinfo{author}{\bibfnamefont{T.}~\bibnamefont{Ohta}} \bibnamefont{et~al.},
  \bibinfo{journal}{Phys.\ Rev.\ Lett.} \textbf{\bibinfo{volume}{98}}
  (\bibinfo{year}{2007}).

\end{thebibliography}

\end{document}